# Features of adsorbed radioactive chemical elements and their isotopes distribution in iodine air filters AU-1500 at nuclear power plants

I. M. Neklyudov, A. N. Dovbnya, N. P. Dikiy, O. P. Ledenyov, Yu. V. Lyashko

*National Scientific Centre Kharkov Institute of Physics and Technology,
Academicheskaya 1, Kharkov 61108, Ukraine.*

The main aim of research is to investigate the physical features of spatial distribution of the adsorbed radioactive chemical elements and their isotopes in the granular filtering medium in the iodine air filters of the type of *AU-1500* in the forced-exhaust ventilation systems at the nuclear power plant. The $\gamma$-activation analysis method is applied to accurately characterize the distribution of the adsorbed radioactive chemical elements and their isotopes in the granular filtering medium in the *AU-1500* iodine air filter after its long term operation at the nuclear power plant. The typical spectrum of the detected chemical elements and their isotopes in the *AU-1500* iodine air filter, which was exposed to the bremsstrahlung $\gamma$ –quantum irradiation, produced by the accelerating electrons in the tantalum target, are obtained. The spatial distributions of the detected chemical element $^{127}I$ and some other chemical elements and their isotopes in the layer of absorber, which was made of the cylindrical coal granules of the type of *SKT-3*, in the *AU-1500* iodine air filter are also researched. The possible influences by the standing acoustic wave of air pressure in the iodine air filter on the spatial distribution of the chemical elements and their isotopes in the iodine air filter are discussed. The comprehensive analysis of obtained research results on the distribution of the adsorbed chemical elements and their isotopes in the absorber of iodine air filter is performed.

## Introduction

The modern energy generation concept in the World, includes both the research on the new clean energy sources and the further development of the existing energy generation technologies. The energy, generated at the nuclear power stations, is considered as a most important energy source for the economies of leading industrial countries. The new strategy of nuclear energy industry development in Ukraine, includes a wide range of research and development programs towards the $4^{th}$ generation nuclear reactors design, including the sodium cooled fast reactor (*SFR*), lead cooled fast reactor (*LFR*), gas cooled fast reactor (*GFR*), very high temperature reactor (*VHTR*), molten salt reactor (*MSR*), supercritical water-cooled reactor (*SCWR*), which can be used for the various technological applications [1]. The air filtering with the application of the iodine air filter of the type of *AU-1500* prevents a possible radioactive contamination of environment due to the introduction of the process of air filtering from the radioactive chemical elements and their isotopes, which are usually generated during the nuclear reactors operation at nuclear power plants. In this research article, we report the research results on the physical features of distribution of the adsorbed radioactive chemical elements and their isotopes in the granular filtering medium (*GFM*) in the iodine air filters (*IAF*) of the type of *AU-1500* in the heating ventilation and cooling (*HVAC*) systems at the nuclear power plant (*NPP*).

The research on the interaction of the aerosol streams, which transfer the radioactive chemical elements and small dispersive coal dust particles masses, with the granular filtering medium, possessing the adsorption properties, is in the scope of interest of the branches of the physics, studying the properties of the granular substance [2], soft condensed matter [3, 4], transport and structurization of small dispersive coal dust particles precipitations [5], and features of chemical elements adsorption processes [6]. In the case of such complex systems, which are usually used as the air filters at the nuclear power stations, there is a big number of physical phenomena, connected with the aerodynamic and diffusion movement of the radioactive chemical elements and their isotopes together with the small dispersive coal dust masses. These effects frequently include both the self-interactions as well as the interactions with the adsorbing granular filtering medium. As it has been found in our previous researches, the distribution of chemical elements, can be detected by the *gamma activation spectroscopy method* [7]. The allocation of dust masses, measured by the *gamma activation spectroscopy method* [7] and the *gravimetric method* [8], is characterized by the presence of correlated maximums in the granular filtering medium in the adsorber in the *IAF*. It is difficult to explain this fact, going only from the diffusion representations about the mechanism of chemical elements transport or considering its similarity to the process, which occurs in the chromatographic columns, where the concentration maximums appearance in the distribution of transported chemical compounds with various molecular masses is observed. Considering other physical mechanisms, which can influence both the distribution of the small dispersive coal dust particles and the adsorption process of the radioactive chemical elements and their isotopes in the granular filtering medium with the cylindrical coal granules, we paid our attention to the fact that the air stream can generate the forced acoustic oscillations of the pressure $P(r, t)$ and the air movement velocity [14]. The layer of granular adsorbent, in which there may be such oscillatory processes, plays a role of the acoustic resonator, whereas the propagating acoustic wave, which is multiple to its thickness, creates the standing acoustic wave oscillations. The pressure oscillations in the air can have an influence on the transposition of the small dispersive coal dust particles masses, causing their accumulation in the regions, corresponding to the anti nodes of the standing acoustic waves [14]. On the other hand, the same oscillations of air density in the regions of the anti nodes of standing acoustic waves will lead to the increase and decrease of the external (in relation to the adsorbent granules) pressure of air, thereby making possible the forced ventilation of the internal air volumes between the cylindrical coal granules. This process should result in the excessive accumulation of the absorbed radioactive chemical elements and their isotopes in these regions. The oscillatory processes can make an influence on the diffusion flows inside the cylindrical coal granules, resulting in an increase of the contents of the adsorbed radioactive chemical compounds in the regions of granular filtering medium in the *IAF*, where the oscillations reach their maximum amplitudes. The same action by the acoustic oscillations must have place at the adsorption processes, which have place in the small dispersive coal dust particles masses, concentrated in the mentioned parts of absorber in the *IAF*. Thus, the penetration of the radioactive chemical elements and their isotopes inside the small dispersive coal dust particles becomes easy, because of their small sizes in comparison with the particles [14]. These acoustic mechanisms should significantly influence the processes of accumulation of the adsorbed radioactive chemical elements and their isotopes, leading to their redistribution inside the *IAF's* absorber in the form of the maximums, which correlate with the nodes of air pressure oscillations. In the previous research [7] with the *gamma activation spectroscopy method* application, it was found that there are the concentration maximums in the distribution of adsorbed radioactive chemical elements and their isotopes in the *IAF*. In the present research, the more detailed improved research data on the distribution of the adsorbed radioactive chemical elements and their isotopes in the *IAF* at the *NPP* during long term operation are presented. The possible physical mechanisms, resulting in a non-uniform distribution of the adsorbed radioactive chemical elements and their isotopes in the *IAF* at the *NPP* are also discussed. The presented highly innovative researches are important, because they can improve our understanding on the nature of the physical-chemical processes in the granular filtering mediums, which are exposed to the air-dust aerosol streams action, helping us to improve the advanced *IAF's* design for the air filtering applications at the *NPP*.

## Experimental setup and measurements methodology

The gamma activation spectroscopy method was applied to research the distribution of the adsorbed radioactive chemical elements and their isotopes in the granular filtering medium of absorber in the *AU-1500 IAF* after many years of the *IAF's* operation at the *Zaporozhskaya* nuclear power plant (*NPP*) in Ukraine. The *IAF* has the following technical characteristics: *1)* the diameter of the *IAF* is about *1 m*; the thickness of layer of cylindrical coal granules is equal to *0,3 m*. The absorber in the *IAF* is made of the cylindrical coal granules of the type of *SKT-3* with the granule's diameter of *2 mm* and the granule's length of *3,2 mm*. During the experimental research, the absorbers were opened with the special automatic experimental setup, then the radioactive probes were taken in the three places (at one place at the center and at the two places at the periphery) in the researched *IAF*. The obtained radioactive probes were extracted at the various depths in the absorber and divided on the samples with the weights of *5, 6, 7 g* with the purpose to perform the chemical composition analysis during their exposition to the electron irradiation at the charged particles accelerator. It is necessary to point out that the macro- and micro- element composition analysis of the adsorbent of the type of *SKT-3* together with its small dispersive coal dust particles fraction was performed by the means of the *gamma activation analysis* at the electron accelerator of the model of *SRC "Accelerator"* at the *NSC KIPT* in Kharkov, Ukraine. The samples were exposed to the irradiation by the bremsstrahlung gamma quantum from the *Tantalum* converter with the thickness of *2 mm*. The energy of electrons was *22 MeV*. The current of electron beam was *700 μA*. The time of exposition to the irradiation was *10 hours*. The activity of the irradiated samples was measured by the *Germanium-Lithium* detector with the volume of *50 cm³* and the energy resolution of *2,8 keV* at the line of *1333 keV*. The absolute contents of the adsorbed radioactive chemical elements and their isotopes were determined in relation to the etalons, which were exposed to the irradiation at the same time. The researches on the determination of fractional composition and the separation of the small dispersive coal dust particles from the cylindrical coal granules were also conducted. It was found that the small dispersive coal dust particles mass share near the input surface of the absorber in the *IAF* reached the value of *12 %*. In the process of measurements, it was established that the accumulated quantity of the small dispersive coal dust particles at the distances far away from the absorber's input surface decreased sharply. The small dispersive coal dust particles mass shares were different at the centre and on the periphery in the absorber in the *IAF*. It makes sense to note that the quantity of the small dispersive coal dust particles was slightly smaller near to the absorber's walls, comparing to the quantity of the small dispersive coal dust particles at the centre of the absorber in the *IAF*.

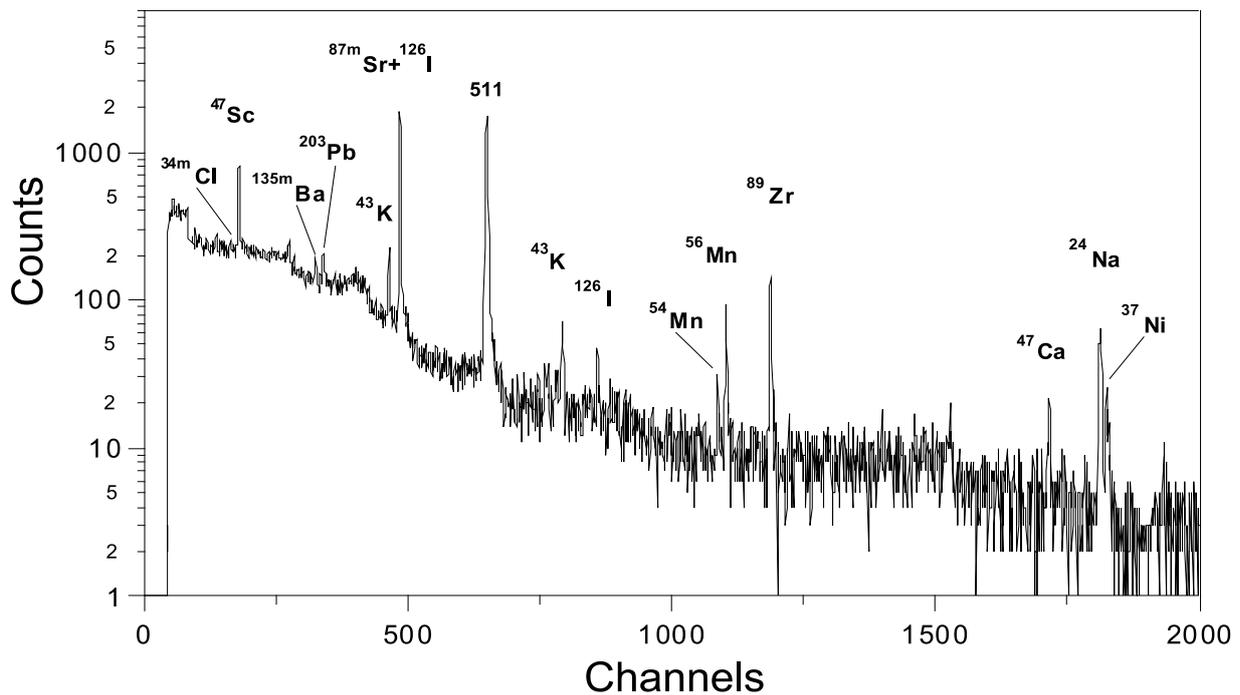

*Fig. 1. Typical spectrum of detected radioactive chemical elements and their isotopes in iodine air filter, exposed to irradiation by electrons at electron accelerator.*

## Measurements results

In the spectroscopy experiment, the induced activities of the isotopes in the following reactions: $^{56}Mn\ (\gamma, n)\ ^{55}Mn$, $^{55}Mn\ (\gamma, n)\ ^{54}Mn$, $^{127}I\ (\gamma, n)\ ^{126}I$, $^{35}Cl\ (\gamma, \gamma)\ ^{35m}Cl$, $^{90}Zr\ (\gamma, n)\ ^{89}Zr$, $^{88}Sr\ (\gamma, n)\ ^{87m}Sr$, $^{48}Ca\ (\gamma, n)\ ^{47}Ca$, $Ca\ (\gamma, p)\ ^{43}K$, $^{23}Na\ (n, \gamma)\ ^{24}Na$, $^{140}Ce\ (\gamma, n)\ ^{139}Ce$, $^{204}Pb\ (\gamma, n)\ ^{203}Pb$, $^{135}Ba\ (\gamma, \gamma)\ ^{135m}Ba$, were measured. The typical spectrum of the induced activity for the probe, taken in the core of the *IAF*, is presented in Fig. 1.

The researched *IAFs* are mainly intended for the chemical adsorption of the radioactive *Iodine (I)* at the *NPP*. Therefore, it was in the scope of our research interest to find out the quantitative content of the radioactive *Iodine* per a mass unit of the adsorbent; and its distribution, depending on the depth of the extracted probe. The measured distribution of $^{127}I$ is shown in Fig. 2. The probes were extracted on the periphery of the *IAF*, and the direction of filtrated air flow as well as the exact position of probe extraction are shown in Fig. 2. The average specific content of small dispersive coal dust particles fraction at the depth of an adsorbent did not exceed *0,5 %*.

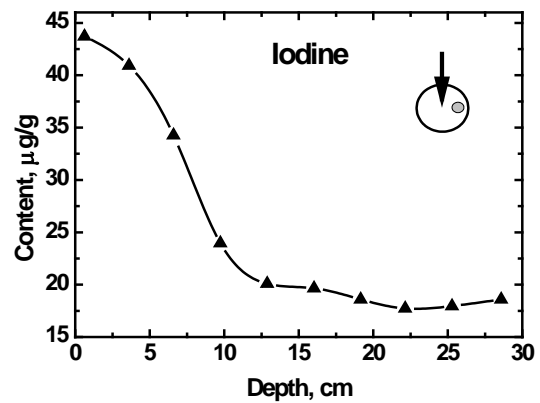

*Fig. 2. Distribution of specific content of Iodine ($^{127}I$) in absorber of iodine air filter at NPP.*

The fission of *Uranium* atoms in a nuclear reactor is accompanied by the accumulation of radioactive isotopes of the *Iodine* ($^{131, 132, 133, 134, 135}I$). It is a well known fact that the half-life decay period for the isotopes of *Iodine* is *8,1 days*; in the case of $^{131}I$; it is *20,9 hours* in the case of $^{133}I$; it is *6,7 hours* in the case of $^{135}I$. Getting into the *IAF*, the *Iodine* isotopes are captured in the *IAF*, where the physical process of their decay takes place. The stable isotope $^{127}I$ is accumulated and preserved in the *IAF* at the moment of experimental measurements. As it can be seen in Fig. 2, the $^{127}I$ is mainly accumulated in the frontal region of the *IAF*, i.e. there is the intensive adsorption of $^{127}I$ as soon as it comes to the *IAF*. The specific content of *Iodine* in close proximity to the input surface of absorber in the *IAF* is *43 μg/g*. Let us note that, according to our measurements results, the initial specific content of *Iodine* in the cylindrical coal granules in an absorber doesn't exceed *5 μg/g*. Some part of decay products of the *Iodine* isotopes could be detected in the reactions $^{135}Ba\,(\gamma,\gamma)\,^{135m}Ba$ and $^{133}Cs\,(\gamma,n)\,^{132}Cs$.

The Figs. 3 and 4 present a relative distribution of the light-weight alkaline chemical elements such as the *Sodium (Na)* and *Kalium (K)* (the $1^{st}$ group of a periodic table of the chemical elements), which were obtained in our measurements.

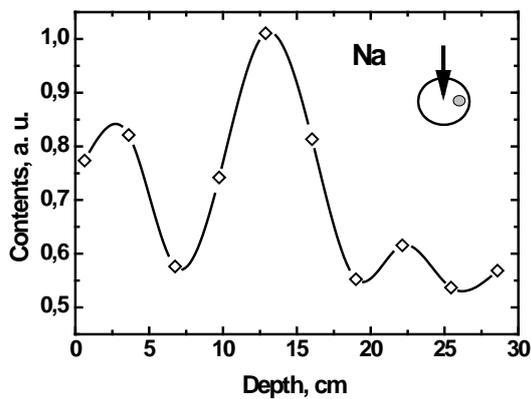

*Fig. 3. Distribution of specific content of Sodium (Na) in absorber of iodine air filter at NPP.*

It is visible that the radioactive chemical elements such as the *Sodium (Na)* and *Kalium (K)* are non-homogeneously distributed in the absorber in the *IAF* and their dependences are well correlated along the length of the *IAF*.

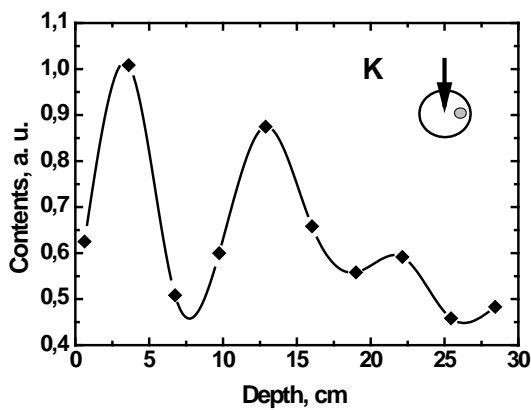

*Fig. 4. Distribution of specific content of Kalium (K) in absorber of iodine air filter at NPP.*

The distributions of radioactive chemical elements from the $2^{nd}$ group from a periodic table of the chemical elements such as the light-weight *Strontium (Sr)* and heavy-weight *Barium (Ba)* are shown in Figs. 5 and 6. These distributions have a high degree of similarity.

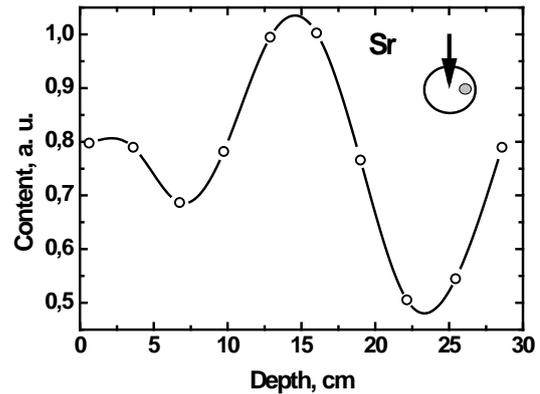

*Fig. 5. Distribution of specific content of Strontium (Sr) in absorber of iodine air filter at NPP.*

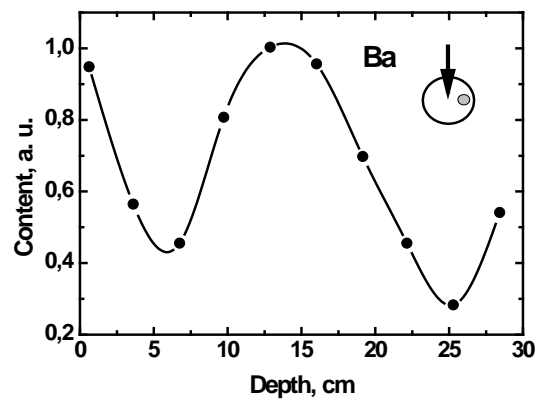

*Fig. 6. Distribution of specific content of Barium (Ba) in absorber of iodine air filter at NPP.*

The lightest chemical element from the $3^{rd}$ group such as the *Scandium (Sc)* is distributed a little differently, but there is a main maximum in the center of absorber in the *IAF* as shown in Fig. 7.

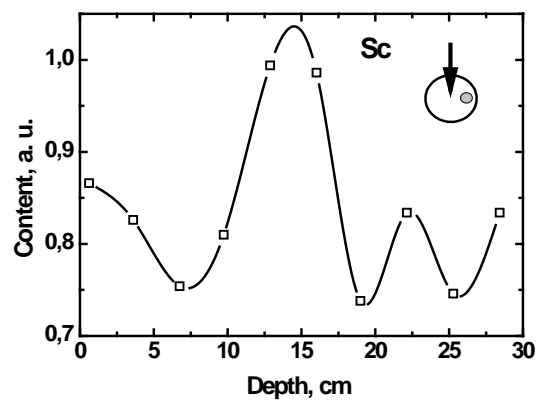

*Fig. 7. Distribution of specific content of Scandium (Sc) in absorber of iodine air filter at NPP.*

The distributions of radioactive chemical elements from the $4^{th}$ group of a periodic table of the chemical elements such as the *Zirconium (Zr)* is given in Fig. 8.

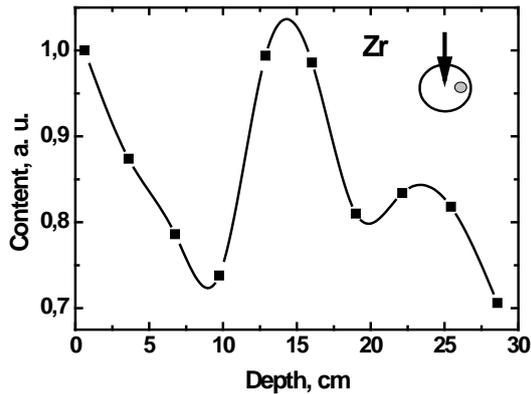

***Fig. 8.*** *Distribution of specific content of Zirconium (Zr) in absorber of iodine air filter at NPP.*

The most light-weighted radioactive chemical element from the $7^{th}$ group of a periodic table of the chemical elements such as the *Manganese (Mn)* has the distribution with the three maximums in Fig. 9.

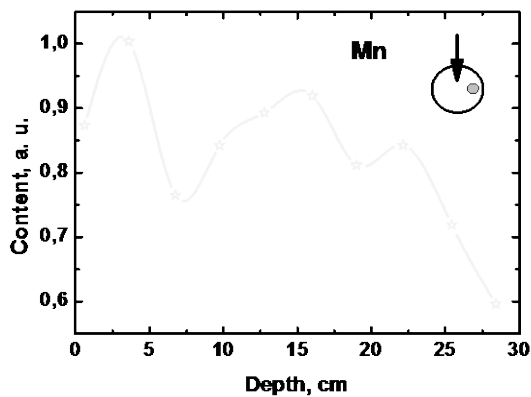

***Fig. 9.*** *Distribution of specific content of Manganese (Mn) in absorber of iodine air filter at NPP.*

In Fig. 10, the distribution of the *Chlorine (Cl)* with the smaller mass, comparing to the *Iodine* mass, from the $17^{th}$ group of a periodic table of the chemical elements is depicted. It is necessary to comment that the distribution of the *Chlorine (Cl)* in Fig. 10 is absolutely different from the distribution of the *Iodine ($^{127}I$)* in Fig. 2.

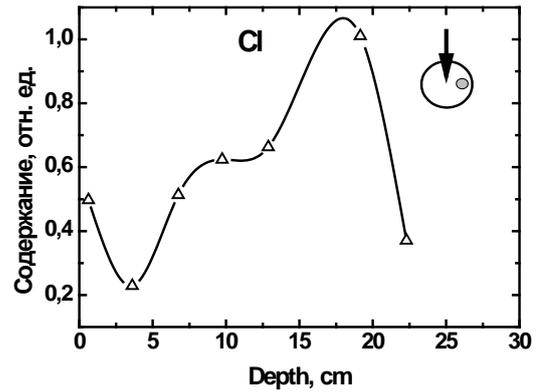

***Fig. 10.*** *Distribution of specific content of Chlorine (Cl) in absorber of iodine air filter at NPP.*

We would like to show the distribution of the relative content of the *Strontium (Sr)* at other places in the absorber in the *IAF*. In Fig. 11, the distribution of the *Strontium (Sr)* in the probe, extracted at the periphery region of absorber in the *IAF*.

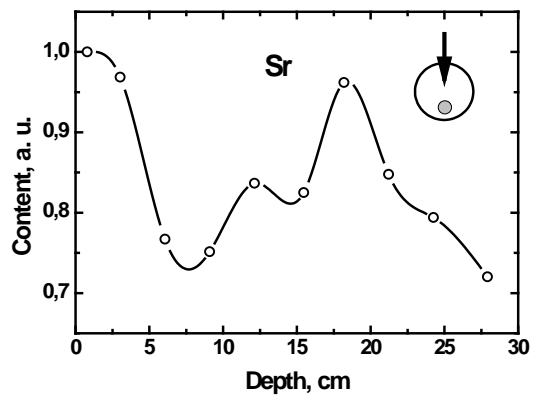

***Fig. 11.*** *Distribution of specific content of Strontium at periphery of absorber in IAF at NPP*

In Fig. 12 the distribution of the *Strontium (Sr)* in the probe, extracted at the center of absorber in the *IAF* is shown.

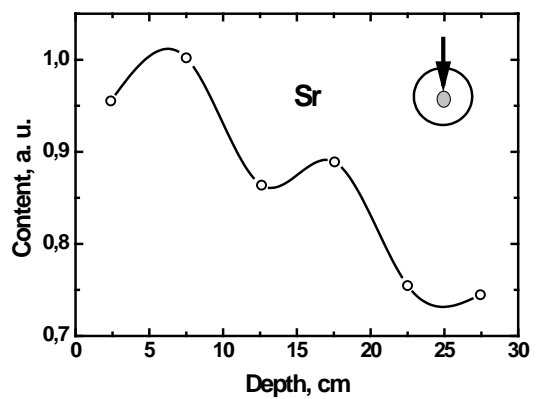

***Fig. 12.*** *Distribution of specific content of Strontium at center of absorber in IAF at NPP*

On the upper net of the *IAF*, the thick and dense layer of the small dispersive coal dust particles, which was not captured by the aerosol filters, was registered. It was found that the direct inflow of the aerosols into the *IAF* from the space, which is in close proximity to the nuclear reactor, worsens the *IAF's* aerodynamic performance. As it was shown during the research with the application of a model of the *IAF*, the presence of the small dispersive dust particles layer on the upper grid increases the absorber's aerodynamic resistance at the rate of *20 %*. The typical spectrum of the detected radioactive chemical elements and their isotopes, which were precipitated at the grid of the *AU-1500* iodine air filter, is presented in Fig. 13. As it can be seen, the presence of $^{137}Cs$, $^{60}Co$, $^{110m}Ag$, $^{54}Mn$ isotopes corresponds to a typical composition of the isotopes, which can be normally detected in the radioactive aerosols at the operating nuclear reactor. The small dispersive dust particles layer on the upper grid of the *IAF* represents the densely packed atmospheric dust, which is mixed with the operating nuclear reactor dust at the *NPP*. The dust layer was indissoluble, because of presence of the oils and moisture. The moisture mass fraction was *5 %*. The particles with the size of about *1 μm* represented a biggest share of particles in the small dispersive dust particles layer, according to the completed microscopy analysis.

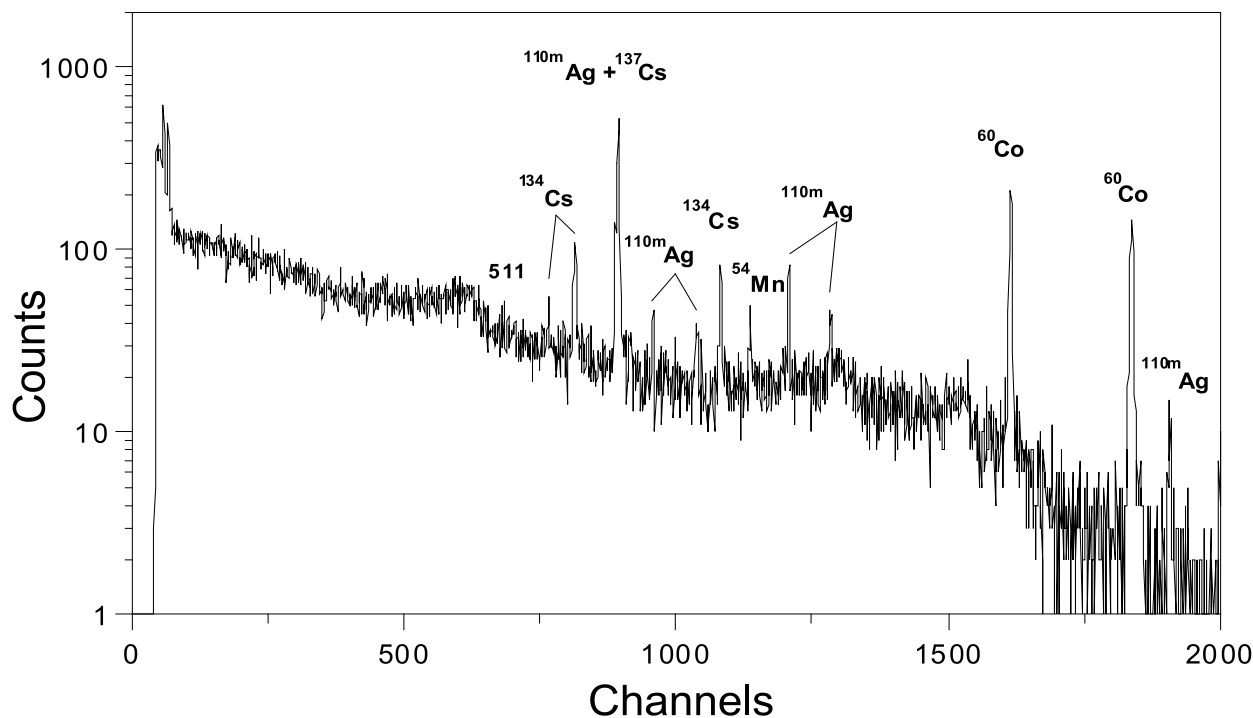

*Fig. 13. Typical spectrum of detected radioactive chemical elements and their isotopes on the grid of iodine air filter, exposed to irradiation by electrons at electron accelerator.*

### Discussion on experimental results

As we explained in the introduction, the distribution and accumulation processes of the adsorbed radioactive chemical elements and their isotopes in the granular filtering medium of absorber in the *AU-1500 IAF* strongly depends on both the physical features of the transportation processes of the small dispersive coal dust particles of the micro- and sub micro-sizes [8-14], as well as the action by the acoustic oscillatory processes, connected with the acoustic oscillations generation and propagation in the air in the granular filtering medium of absorber in the *AU-1500 IAF* [14]. The generating standing acoustic wave oscillations of the first harmonics, which represent the half-wave standing acoustic wave oscillations, described by the dependence between the wavelength and the thickness of granular filtering medium layer: $\lambda_1 = 2L$, and by the next harmonic's dependence: $\lambda_2 = 2L/3$, have the biggest amplitudes. These standing wave oscillations are important for the process of excessive accumulation of the small dispersive coal dust particles, penetrating deeply into the absorber, which takes place in the anti-nodes positions along the length of the *IAF* [14-16]. The pressure maximums in the standing wave oscillations are positioned: *1)* in the case of the first harmonic, it is positioned on the distance of *15 cm*, and *2)* in the case of the second odd harmonic, they are located on the distances of *5, 15, 25 cm* from the input surface of absorber in the *IAF*. The presence of maximums in the distribution of the small dispersive coal dust particles masses has been detected in the experiments with the vertical model of the *IAF* in [8] and the horizontal model of the *IAF* in [12]. According to the research in [9], the maximums of accumulation of the adsorbed radioactive chemical elements and their isotopes in the granular filtering medium of absorber in the *AU-1500 IAF* must be observed in these positions. On the one side, the alternating-sign pressure, appearing in the positions of the anti-nodes of standing wave oscillations, results in an origination of the processes of accelerating penetration of the radioactive chemical elements and their isotopes together with the air into the cylindrical coal granules. From other side, it is a well known fact that such oscillations can intensify the diffusion processes, occurring in the cylindrical coal granules [14].

As it is shown in the experimental data in Fig. 2, the *Iodine* is well adsorbed in the *IAF* at the initial moment of its penetration into the *IAF*. The main share of $^{127}I$ is absorbed by the granular filtering layer with the thickness of *10 cm*. The spatial distribution of *Iodine* has a decreasing diffusion-type dependence [8] on the distance from the source. However, nevertheless, the concentration maximum is clearly visible on the curve of characteristic dependence. This maximum is positioned in the point, where there are the anti-nodes for the two harmonics of the pressure waves in the air in the *IAF*. Therefore, the additional absorption is connected with either: the excessive accumulation of the small dispersive coal dust particles or the influence by the acoustic waves on the adsorption of chemical elements and their isotopes in the granular filtering medium of absorber in the *AU-1500 IAF*. In Fig. 2, there is also a small next maximum in the position near to the absorber's output, which can be related to the action by the acoustic oscillations of second harmonic, which is slightly shifted by the air stream.

The *Alkaline metals* such as the *Sodium* and *Potassium* have the clearly expressed distributions with the maximums, which are located at the forward fronts of anti-nodes, i.e. they are promptly absorbed at the initial stage of filtering process at the moment of entrance to the region of anti-nodes, generated by the acoustic waves. Therefore, their maximums are shifted to the position of source, i.e. to the absorber's input region in the *IAF*.

Moreover, it is necessary to consider the two possibilities: *1)* the cross replacement by the chemical elements and their isotopes in the course of the absorption or *2)* the presence of co-operative absorption in the *IAF*. Thus, for example, the distributions of the radioactive *Strontium* and *Barium* from the $2^{nd}$ group of chemical elements, are characterized by the maximums of absorption (see Figs. 5 and 6), which are similar to the maximum of absorption of *Iodine* (see Fig. 2). Of course, we don't take to the consideration the giant absorption maximum of *Iodine* near to the absorber's input surface in the *IAF* in Fig. 2. Whereas, for example, the *Potassium* (see Fig. 4) was forced to move deeply into the absorber out of absorber's sub-surface region in the *IAF*. It is a well known fact that the *Barium* can easily create the oxides and nitrides. Also, the *Barium* can easily form the $Ba(IO_3)_2$ and the $BaI_2$ chemical compounds, which can have the influences on its distribution in the absorber in the *IAF*. The radiation activity results in the creation of the *Acetic acid* from the *Methane*, containing in the air (the oxidizing of $CH_3$ by the *Hydroxyl OH*):

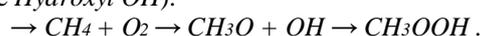
$\rightarrow CH_4 + O_2 \rightarrow CH_3O + OH \rightarrow CH_3OOH$.

The *Acetic acid* can intensively react with the chemical elements of the $2^{nd}$ group.

The *Scandium*, which is a chemical element from the $3^{rd}$ group, has a central maximum of distribution in the region of joint maximums of the both harmonics at the distance *L = 15 cm* from the absorbers input surface in the *IAF*. The *Sc* is accumulated on the forward front of the $1^{st}$ maximum of the $2^{nd}$ harmonic of the standing acoustic wave oscillations at the absorber's input surface; and it is also accumulated on the backward front near to an absorber's output surface in the *IAF*.

The *Zirconium* distribution accompanies the *Iodine* distribution; however in the places with the small concentration of $^{127}I$ at the absorber's output surface, it is concentrated in the region of the $3^{rd}$ maximum of the second harmonic of the standing acoustic wave oscillations at the distance *L = 25 cm* from the absorbers input surface in the *IAF*.

The *Manganese*, which is a chemical element of the $7^{th}$ group, is inclined to occupy the forward fronts of maximums of acoustic oscillations, but, at the center, it occupies the position at the distance *L = 15 cm* from the absorbers input surface in the *IAF* (see Fig. 9).

The *Chlorine*, which is a chemical element from the $17^{th}$ group, occupies the back front of the first maximum of the $2^{nd}$ harmonic of the acoustic oscillations and the back front of the second maximum of the $2^{nd}$ harmonic of the acoustic oscillations, i.e. there is its replacement in the region, where there are no the concentration maximums of other chemical elements (see Fig. 10).

In Figs. 11 and 12, the physical features of distribution of the *Strontium (Sr)* at the different places of probes extraction are shown. It is visible that the air streams can flow non-uniformly, resulting in some non-significant changes of distributions of the adsorbed radioactive chemical elements in the core of the *IAF*. At center of the *IAF*, the *Sr* has a maximum of accumulation near to the absorber's input surface, whereas at the periphery it accumulates in the region at the distance *L = (17...18) cm* from the absorbers input surface in the *IAF*.

In the cases of other chemical elements with the distribution dependences, which are not shown in our research paper, it is necessary to point out that the Calcium has its accumulation maximum at the distance *L = 15 cm*.

Let us stress that the *Ba, Sr, Sc, Zr, Ca* chemical elements can have a positive impact on the absorption capacity of the cylindrical coal granules of the type of *SKT-3*, forming the chemically stable compounds with the *Iodine*.

The almost constant concentrations of the *Ni, Ce, Pb* absorbed chemical elements along the length of the *IAF* was observed. We would like to note that the $PbI_2$ compound is one of the most stable *Iodine's* compounds.

Let us note that the simultaneous consideration of both the diffusion mechanism and the acoustic mechanism in our research on the processes of the accumulation and distribution of the adsorbed radioactive chemical elements and their isotopes in the granular filtering medium of an absorber allows us to explain the obtained complex dependences of the distribution of the adsorbed chemical elements and their isotopes in the granular filtering medium of an absorber in the *AU-1500 IAF* at the *NPP*.

**Conclusion**

The main purpose of our research was to investigate the physical features of distribution of the adsorbed radioactive chemical elements and their isotopes in the granular filtering medium in the iodine air filters of the type of *AU-1500* in the heating ventilation and cooling systems at the nuclear power plant. In our research, the $\gamma$-activation spectroscopy analysis method was applied to accurately characterize the distribution of the adsorbed radioactive chemical elements and their isotopes in the granular filtering medium in the *AU-1500* iodine air filter after its long term operation at the nuclear power plant. We obtained the typical spectrum of the detected chemical elements and their isotopes in the *AU-1500* iodine air filter. The spatial distributions of the detected chemical element $^{127}I$ and some other chemical elements and their isotopes in the layer of absorber, which was made of the cylindrical coal granules of the type of *SKT-3*, in the *AU-1500* iodine air filter, were also comprehensively researched. The possible influences by the changing aerodynamic resistance of the iodine air filter on the spatial distribution of the radioactive chemical elements and their isotopes in the iodine air filter were discussed. In addition, the comprehensive analysis on the obtained research results on the distribution of the adsorbed radioactive chemical elements and their isotopes in the absorber of iodine air filter was performed, the criteria of iodine air filter effective operation were formulated and the possible iodine air filter design improvements were proposed.

Authors are very grateful to a group of distinguished scientists, led by *Boris E. Paton*, at the *National Academy of Sciences in Ukraine (NASU)* for the numerous encouraging scientific discussions on the reported experimental research results.

This innovative research is completed in the frames of the nuclear science and technology fundamental research program, facilitating the environment protection from the radioactive contamination, at the *National Scientific Centre Kharkov Institute of Physics and Technology (NSC KIPT)* in Kharkov in Ukraine.

The research is funded by the *National Academy of Sciences in Ukraine (NASU)*.

This research article was published in the *Problems of Atomic Science and Technology* (*VANT*) in [17] in 2013.

*Corresponding author: ledenyov@kipt.kharkov.ua